\documentclass[12pt]{article}
\usepackage{amssymb,amsmath,mathrsfs}
\usepackage{hyperref}
\usepackage{graphicx}
\usepackage{color}
\usepackage[OT2,OT1]{fontenc}
\usepackage{upquote}
\usepackage{authblk}
\usepackage{mathtools}

\begin{document}

\title{\Large\bf{Carbon dioxide retrieval of Argus 1000 space data by using GENSPECT line-by-line radiative transfer model}}

\bigskip
\author[1, 2, 3]{\small R. K. Jagpal}
\author[1, 2, 3, 4]{\small R. Siddiqui}
\author[1, 2, 3]{\small S. M. Abrarov}
\author[1, 4]{\small B. M. Quine}

\affil[1]{\scriptsize Dept. Physics and Astronomy, York University, 4700 Keele St., Toronto, Canada, M3J 1P3 \normalsize}
\affil[2]{\scriptsize Epic College of Technology, 5670 McAdam Rd., Mississauga, Canada, L4Z 1T2 \normalsize}
\affil[3]{\scriptsize Epic Climate Green (ECG) Inc., 23 Westmore Dr., Unit 310, Toronto, Canada, M9V 3Y7 \normalsize}
\affil[4]{\scriptsize Dept. Earth and Space Science and Engineering, York University, 4700 Keele St., Canada, M3J 1P3 \normalsize}

\date{October 16, 2019}
\maketitle

\begin{abstract}
The micro-spectrometer Argus 1000 being in space continuously monitors the sources and sinks of the trace gases. It is commonly believed that among other gases $\text{CO}_\text{2}$ is the major contributor causing the greenhouse effect. Argus 1000 along its orbit gathers the valuable spectral data that can be analyzed and retrieved. In this paper we present the retrieval of $\text{CO}_\text{2}$ gas in the near infrared window $1580$ to $1620$ nm by using line-by-line code GENSPECT. The retrieved Argus 1000 space data taken over British Columbia on May 31, 2010 indicates an enhancement of $\text{CO}_\text{2}$ by about 30\%.
\vspace{0.25cm}
\\
\noindent {\bf Keywords:} remote sensing; IR detector; carbon dioxide retrieval; micro-spectrometer; line-by-line radiative transfer; atmospheric greenhouse gases
\vspace{0.25cm}
\end{abstract}

\section{Introduction}

Radiative transfer models provide the synthetic radiances as if they could be measured by the sensor for a specified state of the atmosphere. The proper application of a model is essential in any retrieval process. Considering the Earth or an extraterrestrial planetary atmosphere, one typically has to take into consideration the radiation field that describes the angular and/or spatial distribution of the emission and absorption in the Earth atmosphere. One of the most efficient ways to represent accurately atmospheric variation with height is to divide the atmosphere into a large number of relatively thin homogeneous layers or cells where the required parametric values, assigned to each property of interest in each layer, are equal to the corresponding parametric values in the real atmosphere at each mid-point height of the specific layer (see for example (Quine \& Drummond, 2002)).

A transmission through clouds typically is not considered due to complexities related to different effects leading to very inaccurate prediction (Clough et al., 2006; Siddiqui et al., 2015; Siddiqui et al., 2017a; Siddiqui, 2017b, Siddiqui et al., 2019; Su et al., 2013; Cullather et al., 1997). Therefore, the retrievals are mostly attempted on observed profiles where the presence of macroscopic clouds is visibly absent. As the density of the atmosphere varies as a function of height, the refractive index that depends on wavelength also changes with altitude (Clough et al., 2006; Ciddor, 1996). As a result, the radiation entering the atmosphere at any non-zero zenith angle is continuously refracted. Hence, a radiation can in general follow a curved path until either leaving the atmosphere again or reaching the ground.

It is common to assume that during data retrieval process the response of the spectrometer would match with that of simulation obtained by a radiative transfer model of the Earth atmosphere (Quine \& Drummond, 2002; Clough et al., 2006; Jagpal et al., 2010; Fyfe et al., 2013). The ideal match of the retrieved and observed data is exceptionally rare due to number of reasons including imperfectness of the model and instrument variation on successive measurements conditions (Fyfe et al., 2013). Therefore, it is common to consider the retrieved results acceptable when mismatch in radiance does not exceed few percent, say $< 10$\% (see for example SCIAMACI datasets in (Buchwitz et al., 2004; Buchwitz et al., 2005)).

In this paper we present the correlation of experimental data from Argus 1000 micro-spectrometer orbiting currently around the Earth with retrieved data obtained by the radiative transfer model GENSPECT (Quine \& Drummond, 2002). We apply normal or nearly normal radiation, the nadir view, as this approach can effectively ignore the effects of refraction as the angle of refraction is close to zero.

In should be noted that normal incidence can also help avoid more complex radiative transfer geometry associated with tracking the path of radiation through a spherical atmosphere. Consequently, the nadir view consideration is advantages for relatively more accurate simulation of the atmospheric radiation (Buchwitz et al., 2004; Buchwitz. et al., 2000).

\section{Argus real time observations}
 
Argus 1000 observations are available via CanX-2 telemetry (courtesy of the CanX-2 team) since June 2008. With two months of delay after launching the rocket from Satish Dhawan Space Centre, India, on 28 April 2008 (Sarda et al., 2009; Jagpal, 2011), the Argus 1000 micro-spectrometer started sending the observation data. This delay was related to technical issues such as the satellite communication management and telemetry. Since then few hundred datasets have been obtained, most of them over ocean surfaces. Some of the locations corresponding to very recent dataset profiles along with their respective location in are shown in world map Fig. 1 (Siddiqui et al., 2015; Siddiqui, 2017b, Jagpal, 2011).

\begin{figure}[ht]
\begin{center}
\includegraphics[width=28pc]{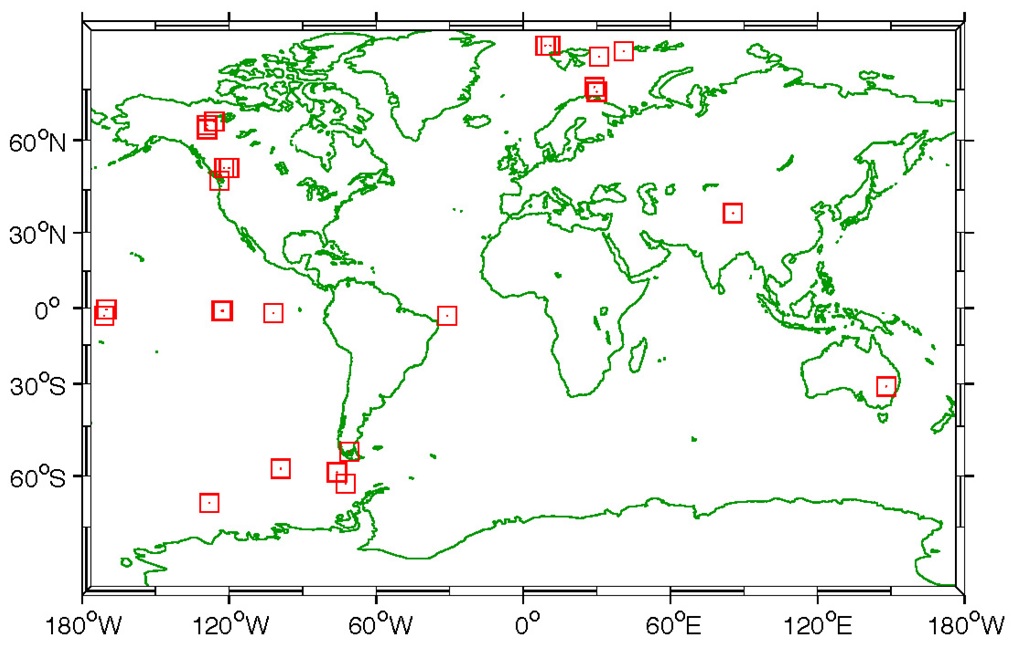}\hspace{2pc}%
\begin{minipage}[b]{28pc}
\vspace{0.3cm}
{\sffamily {\bf{Fig. 1.}} Argus measurement locations for approximately 80 recorded profiles (Siddiqui et al., 2015; Siddiqui, 2017b; Jagpal, 2011).}
\end{minipage}
\end{center}
\end{figure}

Currently, there are only a few satellite instruments in orbit that are designed to measure atmospheric $\text{CO}_\text{2}$. Some of these instruments detect preferably the thermal radiation emitted from the Earth surface rather than the reflected solar radiation. Although in many cases it is advantageous since the measurements are possible not only at day-time but also at night-time, the drawback of such measurements is their lack of sensitivity in the lower troposphere where the strongest signals due to sources and sinks can be expected. Argus 1000 detects both, the thermal radiation as well as the reflected solar radiation, to get the signal originating particularly in the lower troposphere due to industrial pollution and other anthropological activities (Sarda et al., 2009; Jagpal, 2011; Chesser et al., 2012).

In order to convert Argus observations to absolute radiance profiles, three data files, an actual observed profile from Argus in space, a calibration profile using the 1000 W high-calibration etalon lamp, and a background profile for elimination of a background noise, are applied (Jagpal, 2011). Labview software is used as the front end while getting calibration and background profiles. All data analysis is performed with GENSPECT radiative transfer library (Quine \& Drummond, 2002), based on the user-friendly MATLAB programming environment.

\begin{figure}[ht]
\begin{center}
\includegraphics[width=28pc]{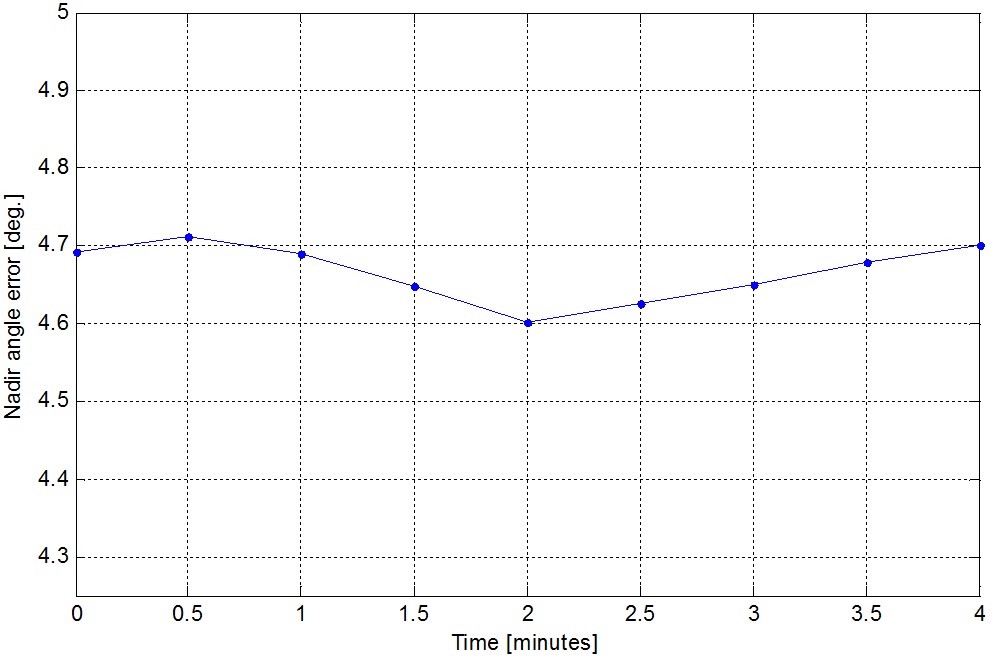}\hspace{2pc}%
\begin{minipage}[b]{28pc}
\vspace{0.3cm}
{\sffamily {\bf{Fig. 2.}} Deviation from nadir view of the Argus 1000 micro-spectrometer obtained from the Canx-2 spacecraft (Siddiqui et al., 2015).}
\end{minipage}
\end{center}
\end{figure}

\section{Geolocations of Argus 1000}

Satellite position and attitude measurements are used for geolocating of Argus 1000 in-orbit micro-spectrometer. In this section we describe the geolocation analysis of the data over British Columbia, Canada. Satellite position is determined applying the SGP4 algorithm with a NORAD two-line element set and satellite attitude measurements as inputs (Jagpal, 2011; Chesser et al., 2012). The NORAD two-line element set is chosen to have an epoch as close as possible to the geolocated observation in order to minimize propagation error. Attitude measurements are obtained in one minute intervals over the duration of requested observations with extra time added before and after the requested observations to take into account the measurement errors (Jagpal, 2011; Chesser et al., 2012). Figure 2 illustrates the nadir angle of approximately $4.7^o$ throughout the observation made on May 31, 2010 between 18:47 and 18:58 UTC. 

The attitude measurements and the spacecraft position estimation are used to determine the footprint of the instrument. The geolocation uncertainties have been estimated to be $33.4$ km along-track and $12.8$ km cross-track (Jagpal, 2011; Chesser et al., 2012). Three sources of geolocation uncertainties are shown in the Table 1.

\begin{table}[ht]
  \begin{center}
		  \begin{tabular}{c|c|c}
       \sffamily{\textbf{Cause of uncertainty}} & \sffamily{\textbf{\begin{tabular}{c}
					Earth central \\
					angle uncertainty \\
					in along-track \\
					direction [mrad]
					\end{tabular}}} & \sffamily{\textbf{\begin{tabular}{c}
					Earth central \\
					angle uncertainty \\
					in cross-track \\
					direction [mrad]
					\end{tabular}}} \\
           \hline
           Attitude determination    & $1.7$ & $1.7$ \\
           NORAD TLE                 & $0.3$ & $0.3$ \\
           Onboard timing system     & $3.2$ & $0$   \\
					 Total                     & $5.2$ & $2.0$ \\
			    \end{tabular}\vspace{0.55cm} \\
			    \sffamily{\begin{tabular}{l}
			     \textbf{Table 1.} Geolocation uncertainties for Argus micro-spectrometer data at \\
					  nadir angle $3.8^o$ (Siddiqui et al., 2015).
		       \end{tabular}}
	  \end{center}
\end{table}

Figure 3 shows the footprint of the observations (made on May 31, 2010 at 18:47 to 18:58), along with the uncertainties corresponding to an ellipse around the observation area.

\begin{figure}[ht]
\begin{center}
\includegraphics[width=28pc]{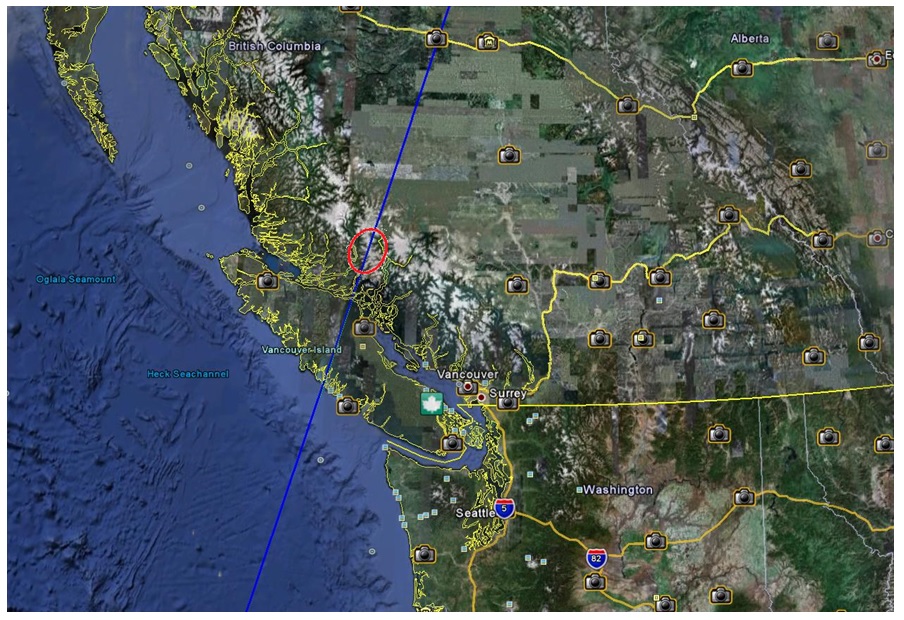}\hspace{2pc}%
\begin{minipage}[b]{28pc}
\vspace{0.3cm}
{\sffamily {\bf{Fig. 3.}} Keyhole Markup Language (KML) output in Google Earth of Argus observations acquired over area in British Colombia, Canada on May 31, 2010 at 15:37:03.02 UTC.}
\end{minipage}
\end{center}
\end{figure}

\section{Retrieval methodology}

The Argus 1000 in the nadir direction measures the atmospheric radiation upwelling vertically from the Earth. While occultation and limb-view techniques typically offer good vertical resolution, the advantages of nadir view are the higher chances to avoid cloud interference, the precise horizontal spatial resolution, the higher potential to retrieve data from the lower atmosphere (below $5$ km), and the more reliable and accurate data retrieval.

For this study, the GENSPECT radiative transfer modeling based on the optimal retrievals methodology is applied (Quine \& Drummond, 2002). This approach is also utilized in the MOPSIG model to output simulated radiance profiles (McKernan et al., 2002). Initially the fraction of emitted radiation arising from different pressure levels (altitudes) at certain temperature fixed for Earth and Sun is computed. We have to take into account the nadir angle for the Argus instrument and Sun zenith angle in the calculation. One of the major factors in simulation is the surface albedo. In optimization process the initial guess value of the surface albedo is taken. Then it is kept changing incrementally to match with the experimental radiance. Finally, the radiative transfer equation is solved numerically to obtain the simulated radiance (Jagpal, 2011). 

If the calculated radiance matches the observed radiance within the typical noise levels of the spectrometer, then the current atmospheric profile (mixing ratio of the gases) is accepted. If the calculated and observed radiances disagree, then the iterated (updated trial) profile is adjusted and the above steps are repeated. Although this method is computationally expensive and requires accurate knowledge of transmittance of each molecular species, it is flexible in optimization process during each stage of the retrieval (Jagpal, 2011).

For a nadir viewing instrument measuring columnar densities of trace species, the atmospheric properties near the surface and surface reflectivity principally determine the nature of radiance profile measured by an instrument at the top of the atmosphere (except for absorption by certain species such as ozone whose concentration are highest far above the Earth’s surface).

\subsection{GENSPECT line-by-line radiative transfer}

GENSPECT computes the absorption and emission of radiation as it passes through gases based on spectral line data and associated physics. The software is able to estimate the spectral effect of changes in near-surface gas composition, and this information can be used in an inversion algorithm to estimate the condition of the atmosphere based on our observations from space. The GENSPECT toolbox has been used previously in the analysis of satellite data from the MOPITT instrument (McKernan et al., 2002; Quine et al., 2005) and will be used as the basis for our development of data-processing routines to retrieve gas amounts in this initiative.

\begin{figure}[ht]
\begin{center}
\includegraphics[width=28pc]{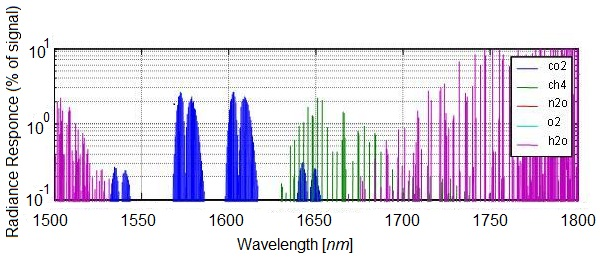}\hspace{2pc}%
\begin{minipage}[b]{28pc}
\vspace{0.3cm}
{\sffamily {\bf{Fig. 4.}} Sensitivity of GENSPECT code showing a 1\% change in concentration of the species shown in the legend. The figure is constructed for atmosphere containing $\text{O}_\text{2}$, $\text{CO}_\text{2}$, $\text{N}_\text{2}\text{O}$, $\text{CH}_\text{4}$, $\text{H}_\text{2}\text{O}$ molecular species at typical volume mixing ratios and isotopes over 0-50 km altitude (Jagpal, 2011).}
\end{minipage}
\end{center}
\end{figure}

GENSPECT is a flexible line-by-line code to study the $\text{CO}_\text{2}$ absorption in the atmosphere. The lines are individually fitted by using Voigt function (Armstrong \& Nicholls, 1972; Abramowitz \& Stegun, 1972; Schreier, 1992; Abrarov et al., 2010; Abrarov \& Quine, 2011; Abrarov et al., 2019) and a polynomial baseline (Quine \& Drummond, 2002). The latest version of the GENSPECT can also utilize a new methodology for line-by-line radiative transfer model based on the spectrally integrated Voigt function that enables rapid computation at reduced spectral resolution (Quine \& Abrarov, 2013). The fitting program uses the HITRAN line strengths, broadening coefficients and lower state energies along with the measured temperature, pressure and path lengths to calculate the mixing ratio for $\text{CO}_\text{2}$ for each isotope (Hill et al., 2016). The tabulated HITRAN line strengths are pre-adjusted for normal isotopic abundances as tabulated, so that the retrieved $\text{CO}_\text{2}$ mixing ratios should be the same for a sample with natural abundance regardless of which line is used. 

Assuming a 150 ms integration time, a 12 mm optical aperture, a typical orbit height of 600 km and a minimum surface albedo of 3\%, we estimate the instrument photon flux over the spectral window to be approximately $5\times10^6$ photons per pixel observation during daylight illumination (giving a shot noise of better than 1:2000). Other factors will further degrade the instrument noise floor, but we compute that the instrument will still yield a column abundance to better than 1\% for $\text{CO}_\text{2}$ and $\text{CH}_\text{4}$. The spectral sensitivity to these gas species is presented in Fig. 4. The retrieval of atmospheric composition data also requires the estimation of surface albedo and atmospheric temperature as well as a data reduction (cloud mask) analysis. A surface resolution of order 1 km will enable the identification of local variation and sources of pollution emission. Coverage will be global but sparse due to the high spatial resolution, specific viewing requirements and instrument operating schedule. Low altitude sensitivity will be maximized through use of solar reflection and a correlation analysis to identify spectral features closely correlated with near-surface abundance.

\subsection{Carbon dioxide retrieval results}

We used the method of optimal retrievals by means of a forward model to output simulated radiance profiles. Previous work (Quine et al., 2005) introduced a non-linear global optimization algorithm to search for a global minimum by iteratively perturbing the full state vector of instrument parameters and trace gas species on the full vertical grid. While this technique produced reasonable results, it required large amounts of computing resources, and time. In order to produce results on a faster time scale we have modified this approach.

GENSPECT introduced an updated retrieval algorithm for the space instrument (Quine \& Drmmond, 2002), based on detailed forward modeling of the atmosphere and instrument (Jagpal, 2011; Quine et al., 2005). Given atmospheric temperature and pressure information, expected trace gas abundances, and some instrument calibration parameters, our model simulates the spectral scans recorded by the instrument during flight. An optimization routine is used to obtain a best fit between the simulated and measured spectra by adjusting the instrument parameters and trace gas amounts. The main advantage of this technique is the incorporation of instrument parameters into the retrieved state vector, which allows the analysis of flight data without pre- and post- calibration data. We summarize the main points of the retrieval algorithm below, with special emphasis on modifications made since 2005 (Quine et al., 2005).

The aim of the instrument forward model is to accurately simulate the true mapping between input radiance and detector response. While it is assumed that the individual instruments are for the most part functionally identical, a few instrument parameters defining unique properties of the each instrument is necessary, and are included in the forward model. Two parameters define the value of the radiance obtained from the instrument are affected by selecting the different values of capacitance setting and exposure time.

A linear relationship between pixel position and wavelength was obtained using calibration are kept in mind here. Initial estimates for the angular position of the instrument are obtained manually by finding the absorption band of the $\text{CO}_\text{2}$, $\text{O}_\text{2}$ and $\text{H}_\text{2}\text{O}$ peaks in the raw data. These instrument parameters are then included in the retrieval, and serve to shift and stretch the data in order to match the simulated spectra.

\begin{figure}[ht]
\begin{center}
\includegraphics[width=28pc]{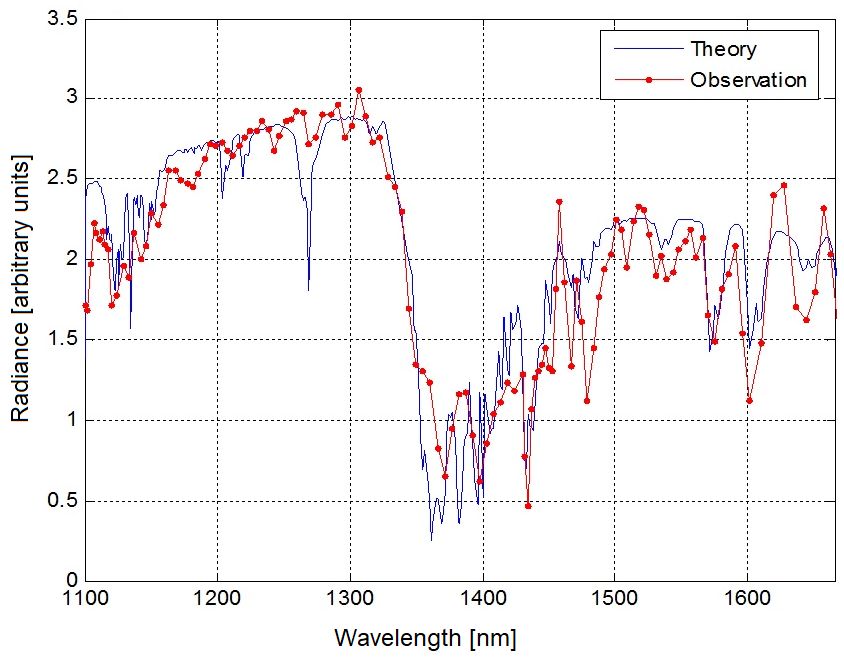}\hspace{2pc}%
\begin{minipage}[b]{28pc}
\vspace{0.3cm}
{\sffamily {\bf{Fig. 5.}} Simulated (GENSPECT) and observed [actual orbital measurement taken on May 31, 2010 [45 degree 25 min N, 75 degree 44 min W] spectrum for Argus 1000 spectrometer. $\text{CO}_\text{2}$ concentration for the forward model is $\sim 30\%$ greater than the standard (384 ppm) concentration.}
\end{minipage}
\end{center}
\end{figure}

A third instrument parameter is used in the construction of the instruments’ slit function. The shape of the slit function, and the relationship between width and center wavelength is based upon laser calibration analysis and spectrometer measurements of the band pass of a sample Argus. A retrieved instrument parameter specifies the width of a 2 sigma truncated Gaussian function convolved with this experimental slit function, accounting for the finite angular width of the focused light passing through the Argus 1000 space sensor.

Following figures 6 and 7 show the results for May 31, 2010. This observation demonstrates a good correlation between the theoretical values and experimental measurements as shown in Fig. 5. As $\text{CO}_\text{2}$ is commonly believed a main contributor to greenhouse effect, its retrieval would be one of our main objectives. As a retrieval tool we applied the GENSPECT code that proved its efficiency and provided reasonable match with experimental observation.

\begin{figure}[ht]
\begin{center}
\includegraphics[width=28pc]{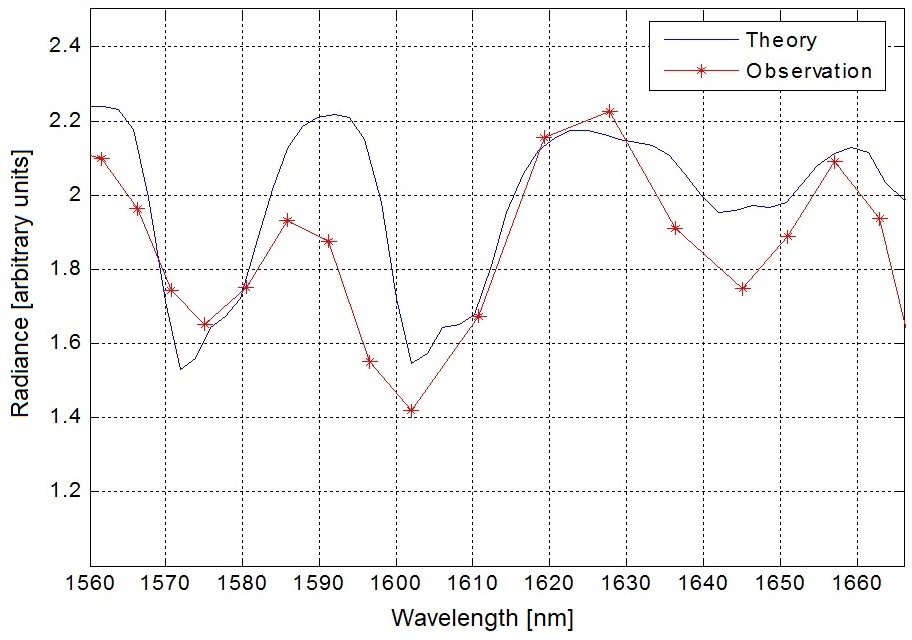}\hspace{2pc}%
\begin{minipage}[b]{28pc}
\vspace{0.3cm}
{\sffamily {\bf{Fig. 6.}} Simulated and experimental radiance for $\text{CO}_\text{2}$ absorption window near 1600 nm.}
\end{minipage}
\end{center}
\end{figure}

Our $\text{CO}_\text{2}$ analysis is based on the fact that in the near infrared region 1580 - 1640 nm it has a spectral window where only this gas contributes for radiance as shown in Fig. 6. Within this spectral window, the difference between observed result and GENSPECT modeled radiance does not exceeds 10\%.

\subsection{Mixing ratio}

Figure 7 shows the contour plot of the relative error.  As we can see, the best match occurs at mixing ratio level corresponding to 1.3. This signifies that $\text{CO}_\text{2}$ concentration enhances up to 30\% over this examined area in British Columbia. The possible reasons for increase of $\text{CO}_\text{2}$ concentration are forest fires (Dhillon, 2010) and/or industrial pollutions (Idso et al., 2001).

The mixing ratio of $\text{CO}_\text{2}$ in the model has been enhanced by a factor of $\sim 1.27$ to obtain a good fit to the radiance profiles observed by Argus. At this time, effects due to atmospheric effects such as cloud and double refraction are not included in the current version of GENSPECT. It is known that scattering by clouds can potentially introduce errors in the retrieved column amount of trace gases by adding uncertainty to the radiation path length. Clouds (and other airborne particles) can absorb or scatter sunlight back to space before it traverses the full atmospheric column, precluding full-column $\text{CO}_\text{2}$ measurements in regions occupied by opaque clouds. The relative error map that was calculated according to the formula
$$
{\Delta _R} = \frac{{\left| {{R_t} - {R_e}} \right|}}{{{R_t}}},
$$
where $R_t$ and $R_e$ ere theoretical GENSPECT and experimental Argus 1000 radiances, respectively. From this figure we can see that mixing ratio factor at $1.27$ illustrates the best match between the theoretical and experimental radiance data indicating the increase of $\text{CO}_\text{2}$ by amount up to 30\%. This value of $\text{CO}_\text{2}$ enhancement is in a good agreement with other results obtained over a polluted area due to anthropogenic activities. For example, Idso et al. reported 43\% enhancement of $\text{CO}_\text{2}$ in Phoenix, Arizona (Idso et al., 2001). Significant seasonal variation of mixing ratio of $\text{CO}_\text{2}$ in London, UK, were reported in the paper (Rigby et al., 2008).

\begin{figure}[ht]
\begin{center}
\includegraphics[width=28pc]{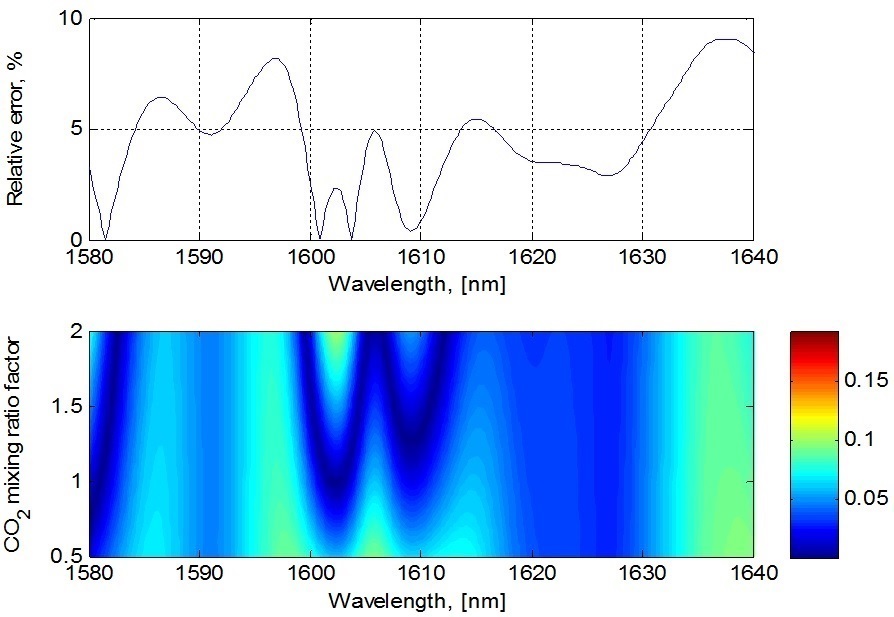}\hspace{2pc}%
\begin{minipage}[b]{28pc}
\vspace{0.3cm}
{\sffamily {\bf{Fig. 7.}} Determination of mixing ratio from Argus: Upper panel is relative error in \% obtained from the comparison of simulated and observed radiances over an area in British Columbia [Solar zenith angle $= 67^o$, nadir angle $=5^o$].}
\end{minipage}
\end{center}
\end{figure}

\section{Conclusion}

The space orbiting Argus 1000 micro-spectrometer continuously monitors the sources and sinks of the trace gases. The retrieved Argus 1000 space data sample taken over British Columbia on May 31, 2010 shows an enhancement of $\text{CO}_\text{2}$ by about 30\%. This value of $\text{CO}_\text{2}$ enhancement is in a good agreement with results taken over a polluted area due to human activities. For many years of heritage in space, Argus 1000 demonstrates that the new generation of the light, small and inexpensive micro-spectrometers can be efficient in global monitoring of $\text{CO}_\text{2}$ gas, mostly responsible for greenhouse effect on the Earth.

\section*{Acknowledgments}

This work is supported by National Research Council Canada, Thoth Technology Inc., York University, Epic College of Technology and Epic Climate Green (ECG) Inc. The CanX-2 space mission is supported by Defense Research and Development Canada (Ottawa), MacDonald Dettwiler and Associates Space Missions, Dynacon Incorporated, Ontario Centers of Excellence, ETech Division, Canadian Space Agency and Radio Amateur Satellite Corporation (AMSAT). The authors wish to thank Prof. Robert E. Zee and his team, University of Toronto Institute for Aerospace Studies, for facilitating the first flight of Argus 1000 spectrometer. We would like to acknowledge significant contribution of Prof. John C. McConnell (late) and Dr. Brian Solheim, York University, to this research project. Comments and constructive discussions from Harshal Gunde, Epic Climate Green (ECG) Inc., are greatly appreciated.

\bigskip

\end{document}